\documentstyle[11pt,epsf]{article}

\newcommand{\lb}{\label}

\newcommand{\bc}{\begin{center}}
\newcommand{\ec}{\end{center}}
\newcommand{\bd}{\begin{displaymath}}
\newcommand{\ed}{\end{displaymath}}
\newcommand{\be}{\begin{equation}}
\newcommand{\ee}{\end{equation}}
\newcommand{\ba}{\begin{array}}
\newcommand{\ea}{\end{array}}
\newcommand{\bt}{\begin{tabular}}
\newcommand{\et}{\end{tabular}}

\newcommand{\ov}{\overline}
\newcommand{\bp}{\begin{picture}}
\newcommand{\ep}{\end{picture}}
\newcommand{\bfi}{\begin{figure}}
\newcommand{\efi}{\end{figure}}

\begin{document}


\title{\Large \bf The Problem of Monopoles in the Standard and Family Replicated Models}

\author{L.Laperashvili${}^{1}$,
H.B.Nielsen${}^{2}$\\[15mm]
\itshape{${}^{1}$ Theory Department,
ITEP, Moscow, Russia}\\[3mm] \itshape{${}^{2}$ The Niels Bohr
Institute, Copenhagen, Denmark }}

\date{}

\maketitle

Talk given by L.V.Laperashvili at the Eleventh Lomonosov
Conference On Elementary Particle Physics, Moscow, Russia, 21-27
August, 2003.

\begin{abstract}
The aim of the present talk is to show that monopoles cannot play
any role in the Standard Model (SM), and in its usual extensions,
up to the Planck scale: $M_{Pl}=1.22\cdot 10^{19}$ GeV, because
they have a huge charge and are completely confined or screened.
The possibility of the extension of the SM with a Family
Replicated Gauge Group (FRGG) symmetry of the type $(SMG)^N =
[SU(3)_c]^N\times [SU(2)_L]^N\times [U(1)_Y]^N$ is briefly
discussed. It was shown that the Abelian monopoles (existing also
in non-Abelian theories) in FRGG model have $N^*$ times smaller
magnetic charge than in the SM, where $N^*=N(N+1)/2$. These
monopoles can appear at the high energies in the FRGG-model and
give additional contributions to the beta--functions of the
renormalisation group equations for the running constants
$\alpha_i(\mu)$, where i=1,2,3 correspond to the U(1), SU(2) and
SU(3) gauge groups of the SM.
\end{abstract}

\vspace{3cm} \footnoterule{\noindent ${}^{1}$ E-mail:
laper@heron.itep.ru\\ ${}^{2}$ E-mail: hbech@alf.nbi.dk}

\newpage

\thispagestyle{empty}

\pagenumbering{arabic}

\newpage

\section{The Problem of Monopoles in the Standard Model}

The gauge symmetry group in the SM is :
\begin{equation}
   SMG = SU(3)_{c(color)}\times SU(2)_{L(left)}\times
U(1)_{Y(hypercharge)},
            \lb{1}
\end{equation}
which describes the present elementary particle physics up to the scale
$\approx 100$ GeV.

The aim of the present talk is to show that monopoles cannot be
seen in the Standard Model and in its usual extensions, known in
the literature, up to the Planck scale [1,2]:
\begin{equation}
     M_{Pl}=1.22\cdot 10^{19}\,{\mbox{GeV}},     \lb{7m}
\end{equation}
because they have a huge magnetic charge and are completely
confined or screened.

Supersymmetry does not help to see monopoles.

Let us consider the "electric" and "magnetic" running constants:
\begin{equation}
    \alpha = \frac{g^2}{4\pi}\quad{\mbox{and}}\quad
    \tilde \alpha = \frac{\tilde g^2}{4\pi},                   \lb{10m}
\end{equation}
where $g$ is the coupling constant,

and $ \tilde g$ is the dual coupling constant.

In QED: $$
   g = e \quad - \quad {\mbox{electric charge}},
$$ $$
   \tilde g = m\quad - \quad {\mbox{magnetic charge}}.
$$
The Renormalization Group Equation (RGE) for monopoles is:
\be
\frac {d(\log \tilde \alpha(t))}{dt} = \beta(\tilde \alpha).
               \lb{16m}
\ee Here $t$ is the evolution variable:
\begin{equation}
         t = \log(\frac{\mu^2}{\mu_R^2}),            \lb{12m}
\end{equation}
where $\mu $ is the energy scale and $\mu_R$ is the
renormalisation point.

The scalar monopole beta--function is taken from the dual scalar
electrodynamics by Coleman and Weinberg [3]:
\be
   \beta(\tilde \alpha) = \frac{\tilde \alpha}{12\pi } +
{(\frac{\tilde \alpha}{4\pi })}^2 + ... = \frac{\tilde
\alpha}{12\pi }( 1 +
 3\frac{\tilde \alpha}{4\pi } + ...).                \lb{1m}
\ee
The last equation shows that the theory of monopoles cannot be
considered perturbatively at least for
\be
         \tilde \alpha > \frac{4\pi}{3}\approx 4.
                                  \lb{2m}
\ee {\bf And this limit is smaller for non--Abelian monopoles. }

Let us consider now the evolution of the SM running fine structure
constants $\alpha_i(t)$, where i=1,2,3 correspond to U(1), SU(2)
and SU(3) gauge groups of the SM.

The usual definition of the SM coupling constants is given in {\bf
the Modified minimal subtraction scheme}($\ov{MS}$):
\be
  \alpha_1 = \frac{5}{3}\alpha_Y,\quad
\ee
\be
  \alpha_Y = \frac{\alpha}{\cos^2\theta_{\ov{MS}}},\quad
\ee
\be
   \alpha_2 = \frac{\alpha}{\sin^2\theta_{\ov{MS}}},\quad
\ee
\be
  \alpha_3 \equiv \alpha_s = \frac {g^2_s}{4\pi},
\ee where $\alpha$ and $\alpha_s$ are the electromagnetic and
SU(3) fine structure constants respectively, $Y$ is the weak
hypercharge, and $\theta_{\ov{MS}}$ is the Weinberg weak angle in
$\ov{MS}$ scheme.

Using RGEs with experimentally established parameters, it is
possible to extrapolate the experimental values of three inverse
running constants $\alpha_Y^{-1}(\mu)$ and $\alpha_i^{-1}(\mu)$
(for i=2,3) from the Electroweak scale to the Planck scale (see
Fig.1).

\begin{figure}[t]
\centerline{\epsfxsize=\textwidth \epsfbox{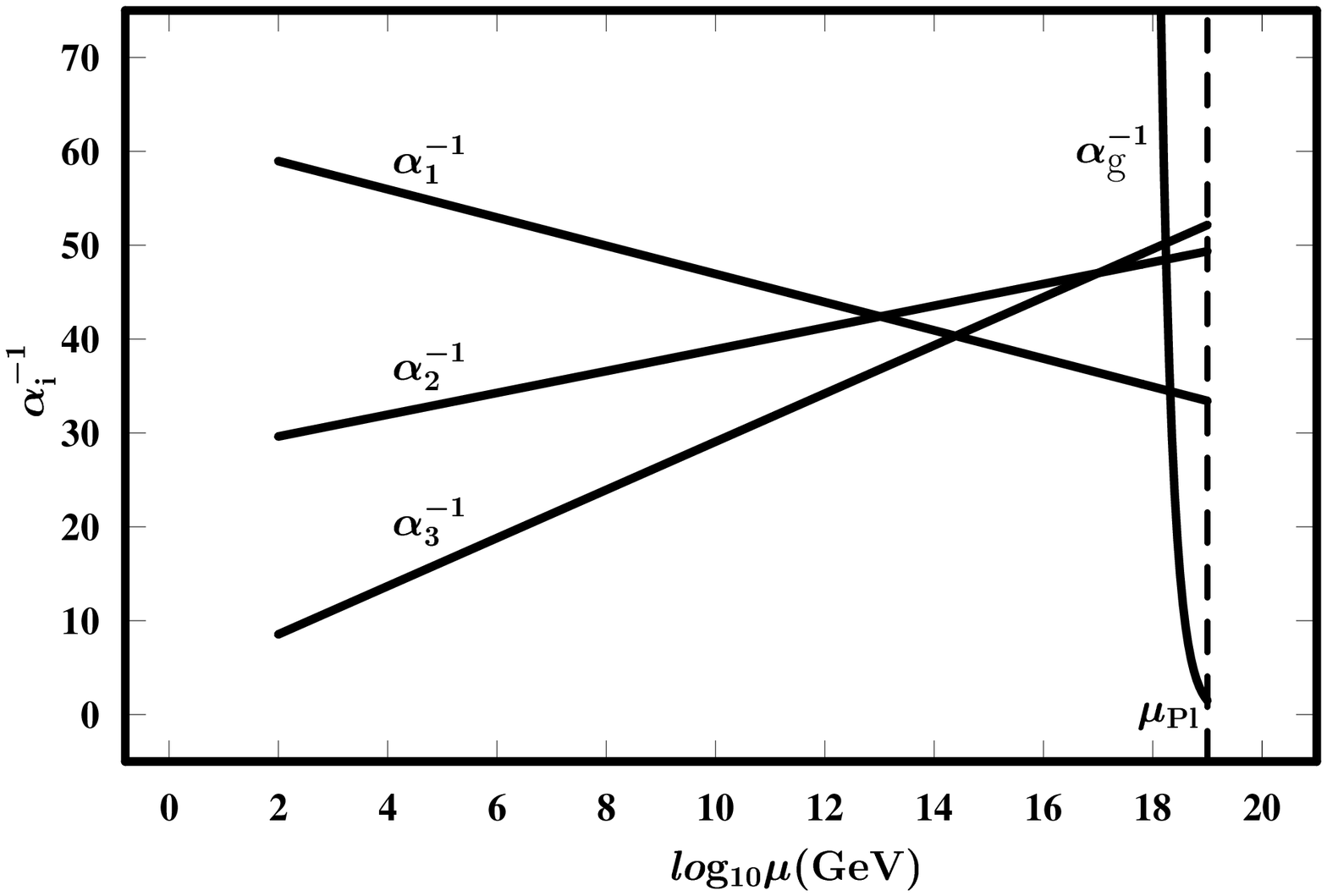}} \caption{}
\end{figure}

Assuming the existence of the Dirac relation for renormalised
charges $g$ and $\tilde g$ [4]:
$$ g\tilde g = 2\pi n, \quad n\in Z,$$
we have for minimal charges n=1 and the following expression:
\be
       \alpha(t) \tilde \alpha(t)= \frac{1}{4}.
\ee
Using this relation, it is easy to estimate (in the simple SM)
the Planck scale value of $\tilde \alpha(\mu_{Pl})$ (minimal for
$U(1)_Y$ gauge group):
\be
 \tilde \alpha(\mu_{Pl}) = \frac{5}{3}\alpha_1^{-1}(\mu_{Pl})/4
           \approx 55.5/4 \approx 14.                   \lb{3m}
\ee This value is really very big compared with our previous
estimate (\ref{2m}) and, of course, with the critical coupling
$\alpha_{crit}\approx 1$, corresponding to the
confinement---deconfinement phase transition in the lattice U(1)
gauge theory.

Clearly we cannot make a perturbation approximation with such a
strong coupling $\tilde{\alpha}$.

It is hard for such monopoles not to be confined.

There is an interesting way out of this problem if one wants to
have the existence of monopoles, namely to extend the SM gauge
group so cleverly that certain selected linear combinations of
charges get bigger electric couplings than the corresponding SM
couplings. That could make the monopoles which, for these certain
linear combinations of charges, couple more weakly and thus have a
better chance of being allowed "to exist".

An example of such an extension of the SM that can impose the
possibility of allowing the existence of free monopoles is just
Family Replicated Gauge Group Model (FRGGM).

\section{Family Replicated Gauge Group as an extension of the Standard Model}

The extension of the Standard Model with the Family Replicated
Gauge Group :
\begin{equation}
G = (SMG)^{N_{fam}} = [SU(3)_c]^{N_{fam}}\times
[SU(2)_L]^{N_{fam}} \times [U(1)_Y]^{N_{fam}}       \lb{2}
\end{equation}
was first suggested in the paper [5] and developed in the book [6]
(see also the review [7]).

Here $N_{fam}$ designates the number of quark and lepton families.

If $\,N_{fam}=3$ (as our theory predicts [6] and experiment
confirms), then the fundamental gauge group G is:
\begin{equation}
G = (SMG)^3 = SMG_{1st\;fam.}\times
SMG_{2nd\;fam.}\times SMG_{3rd\;fam.},
                                        \lb{3}
\end{equation}
or
\begin{equation}
G = (SMG)^3 = {[SU(3)_c]}^3\times {[SU(2)_L]}^3\times
{[U(1)_Y]}^3.
                                        \lb{4}
\end{equation}
A new generalization of our FRGG--model was suggested in papers
[8-10].

The group :
 $$  G_{\mbox{ext}} = (SMG\times U(1)_{B-L})^3 $$
\begin{equation}
 \equiv [SU(3)_c]^3\times [SU(2)_L]^3\times
[U(1)_Y]^3\times [U(1)_{(B-L)}]^3
                                    \lb{6}
\end{equation}
is the fundamental gauge group, which takes right-handed neutrinos
and the see--saw mechanism into account. This extended model can
describe all modern neutrino experiments, giving a reasonable fit
to all the quark-lepton masses and mixing angles in the SM.

The group $G_{\mbox{ext}}$ contains: $$ 3\times 8 = 24 \quad
{\mbox{gluons}},$$ $$3\times 3 = 9 \quad {\mbox{W-bosons}},$$ and
$$3\times 1 + 3\times 1 = 6\quad {\mbox{ Abelian gauge
bosons}}.$$

The gauge group $$ G_{\mbox{ext}} = (SMG\times U(1)_{B-L})^3
$$ undergoes spontaneous breakdown (at some orders of
magnitude below the Planck scale) to the Standard Model Group SMG
which is the diagonal subgroup of the group $ G_{\mbox{ext}}$.

As was shown in the paper [8],  6 different Higgs fields:
$$\omega,\,\,\rho,\,\,W,\,\,T,\,\,\phi_{WS},\,\,\phi_{B-L}$$ break
our FRGG--model to the SM.

The field $\phi_{WS}$ corresponds to the Weinberg--Salam
Electroweak theory. Its vacuum expectation value (VEV) is fixed by
the Fermi constant:
 $$ <\phi_{WS}>\approx 246 \quad{\mbox{GeV}},$$
so that {\bf we have only 5 free parameters -- five remaining VEVs
-- to fit the experiment in the framework of the SM.}

These five adjustable parameters were used with the aim of finding
the best fit to experimental data for all fermion masses and
mixing angles in the SM, and also to explain the neutrino
oscillation experiments.

Finally, we conclude that our theory with the $FRGG$--symmetry is
very successful in describing experiment.

\section{Monopoles in the  Family Replicated Gauge Group Model.}

In theories with the FRGG symmetry the charge of monopoles is
essentially diminished.

Then monopoles can appear near the Planck scale and change the
evolution of the running constants $\alpha_i(t).$

Family replicated gauge groups of type: $$ [SU(N)]^{N_{fam}}$$
lead to the lowering of the magnetic charge of the monopole
belonging to one family:
\be
     \tilde \alpha_{one\,\,family} = \frac{\tilde \alpha}{N_{fam}}.
\ee
For $N_{fam} = 3$, for $[SU(2)]^3$ and $[SU(3)]^3$, we have:
\be
 \tilde \alpha_{one\,\,family}^{(2,3)} = \frac{{\tilde
\alpha}^{(2,3)}}{3}.
\ee
For the family replicated group
$[U(1)]^{N_{fam}}$ we obtain:
\be
    \tilde \alpha_{one\,\,family} = \frac{\tilde \alpha}{N^{*}},
\ee
where
$$
        N^{*} = \frac{1}{2}N_{fam}(N_{fam} + 1).
$$

For $N_{fam} = 3$ and $[U(1)]^3$, we have:
\be
    \tilde \alpha_{one\,\,family}^{(1)} = \frac{{\tilde \alpha}^{(1)}}{6},
\ee
{\bf Six times smaller!}

This result was obtained previously in the paper [11].

According to the FRGGM, at some point $\mu=\mu_G < \mu_{Pl}$ (or
really in a couple of steps) the fundamental group $G\equiv
G_{\mbox{ext}} $ undergoes spontaneous breakdown to its diagonal
subgroup:
\begin{equation}
      G \longrightarrow G_{diag.subgr.} = \{g,g,g || g\in SMG\},
                                                          \lb{9m}
\end{equation}
which is identified with the usual (low--energy) group SMG.

The aim of this investigation is to show that we have the
influence of monopoles with masses:
 \be
 M_{mon} > 10^{14}\,\,{\mbox{GeV}},  \lb{9mon}
\ee if the $G$--group undergoes the breakdown to its diagonal
subgroup (that is, SMG) at
\be
  \mu_G\sim 10^{14}\quad {\mbox{or}} \quad 10^{15}\,\,
  {\mbox{GeV}},                                         \lb{9int}
\ee that is, before the intersection of $\alpha_{2}^{-1}(\mu)$
with $\alpha_{3}^{-1}(\mu)$ at $\mu\approx 10^{16}$ GeV.

In this case, in the region $$\mu_G < \mu < \mu_{Pl}$$ there are
three $SMG\times U(1)_{B-L}$ groups for the three FRGG families,
and we have a lot of new fermions, mass protected or not mass
protected, belonging to usual families or to mirror ones, because
{\bf in the FRGGM the additional 5 Higgs bosons, with their large
VEVs, are responsible for the mass protection of a lot of new
fermions appearing in the region $\mu > \mu_G$.}

In this region we denote the total number of fermions  $N_F$,
which is different to $N_{fam}$.

Also the role of monopoles can be important in the vicinity of the
Planck scale: they can give contributions to the corresponding
beta-functions and change the evolution of $\alpha_i^{-1}(\mu )$.

Here it is necessary to comment: in the FRGG model, near the
Planck scale, monopole charges, together with electric ones, are
sufficiently small, and their $\beta$-functions can be considered
perturbatively:
\begin{equation}
\beta(\alpha) = \beta_{2} (\alpha /4\pi) + \beta_{4}
{(\alpha/4\pi)}^2 + ... \lb{18x}
\end{equation}
\begin{equation}
\beta(\tilde \alpha) = \beta_{2} (\tilde \alpha /4\pi) + \beta_4
{(\tilde \alpha/4\pi)}^2 + ... \lb{19x}
\end{equation}
As was shown in the paper [4], there exists a region when both
running constants $ \alpha $ and $\tilde \alpha $ are
perturbative. Approximately this region is given by the following
inequalities:
\begin{equation}
0.2 \stackrel{<}{\sim }(\alpha, \tilde \alpha)
\stackrel{<}{\sim }1,
                  \lb{22x}
\end{equation}
In this region the two-loop contribution to beta-function is not
larger than 30\% of the one-loop contribution, and the
perturbation theory can be realized in this case.

It is very interesting that the above-mentioned region coincides
with the region of critical couplings for the phase transition
"confinement-deconfinement" obtained in the lattice compact QED:
\be
        \alpha_{crit}^{lat}\approx 0.20\pm 0.015, \lb{24x}
\ee
\be
     \tilde  \alpha_{crit}^{lat}\approx 1.25\pm 0.10, \lb{25x}
\ee obtained in Refs.[12-14], what confirms the idea of Ref.[11]
that at the Planck scale we have the Multiple Critical Point.

\section{The Evolution of Running Fine Structure Constants}

Finally, we obtain the following RGEs:
\begin{equation}
   \frac {d(\alpha_i^{-1}(\mu))}{dt} = \frac{b_i}{4\pi } +
   \frac{N_M^{(i)}}{\alpha_i}\beta^{(m)}(\tilde \alpha_{U(1)}),  \lb{G3m}
\end{equation}
where $b_i$ are given by the following values:
$$
   b_i = (b_1, b_2, b_3) =
$$
\be
( - \frac{4N_F}{3} -\frac{1}{10}N_S,\quad
      \frac{22}{3}N_V - \frac{4N_F}{3} -\frac{1}{6}N_S,\quad
      11 N_V - \frac{4N_F}{3} ).                   \lb{G4m}
\ee The integers $$N_F,\,N_S,\,N_V,\,N_M\, $$ are respectively the
total numbers of fermions, Higgs bosons, vector gauge fields and
scalar monopoles in the FRGGM considered in our theory.

In our FRGG model we have: $$
      N_V = 3,
$$ because we have 3 times more gauge fields $(N_{fam}=3)$,
in comparison with the SM, and one Higgs scalar monopole in each
family.

We have obtained the evolutions of $\alpha_i^{-1}(\mu )$ near the
Planck scale by numerical calculations for: $$\mu_G=10^{14}\,\,
GeV, $$ $$M_{mon}>10^{14}\,\, GeV, $$
$$N_F=18, $$ $$N_S = 6, $$ $$N_M^{(1)} = 6, $$ $$N_M^{(2,3)} = 3. $$

Fig.2  shows the existence of the unification point.

\begin{figure}[t]
\centerline{\epsfxsize=\textwidth \epsfbox{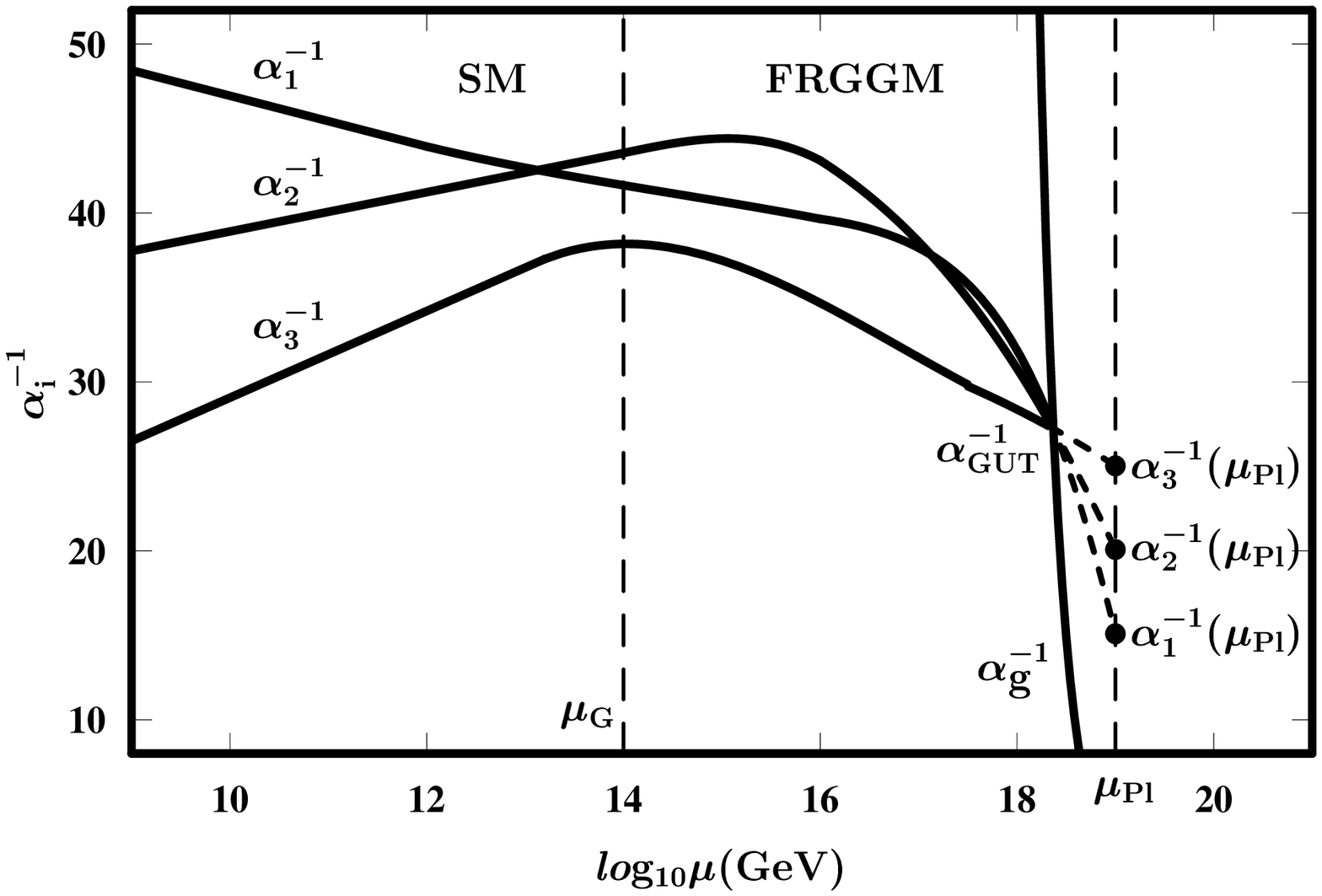}} \caption{}
\end{figure}

In this connection, it is very attractive to include gravity. The
quantity:
\begin{equation}
      \alpha_g = \left(\frac{\mu}{\mu_{Pl}}\right)^2     \lb{2x}
\end{equation}
plays the role of the running "gravitational fine structure
constant" and the evolution of its inverse is presented in Fig.2
together with the evolutions of $\alpha_i^{-1}(\mu)$.

In Fig.2 we see that in the region $\mu > \mu_G$ {\bf a lot of new
fermions and a number of monopoles near the Planck scale change
the one--loop approximation behaviour of $\alpha_i^{-1}(\mu)$
which we had in the SM.}

In the vicinity of the Planck scale these evolutions begin to
decrease, as the Planck scale $\mu = M_{Pl}$ is approached,
implying {\bf the suppression of asymptotic freedom in the
non--Abelian theories.}

Fig.2 gives the following Planck scale values of $\alpha_i$:
\be
   \alpha_1^{-1}(\mu_{Pl})\approx 13,
\ee
\be
   \alpha_2^{-1}(\mu_{Pl})\approx 19,
\ee
\be
  \alpha_3^{-1}(\mu_{Pl})\approx 24.  \lb{G11m}
\ee Fig.2 demonstrates the unification of all gauge interactions,
including gravity (the intersection of $\alpha_g^{-1}$ with $
\alpha_i^{-1}$), at
\be \alpha_{GUT}^{-1}\approx 27 \quad and
\quad x_{GUT}\approx 18.4.
\ee
It is easy to calculate that for
one family we have:
\be \tilde{\alpha}_{GUT, onefam.}=\frac{
\alpha_{GUT, one fam.}^{-1}}{4}=\frac{\alpha_{GUT}^{-1}}{4\cdot
6}\approx \frac{ 27}{ 24}\approx 1.125,
\ee and
\be
\alpha_{GUT, one fam.}\approx 0.22,
\ee what means that at the GUT
scale electric and monopole charges are not large and can be
considered perturbatively.

Here we can expect the existence of $$[SU(5)]^3,\quad \quad
{\mbox{or}}\quad \quad[SO(10)]^3, $$ SUSY or not SUSY unification.

If SUSY, then we have superparticles of masses:
\be
 M\approx 10^{18.4}\quad{\mbox{GeV}}.      \lb{G12m}
\ee The scale $\mu_{GUT}=M$, given by Eq.(\ref{G12m}), can be
considered as a SUSY breaking scale.

Considering the predictions of such a theory for the low--energy
physics and cosmology, maybe in future we shall be able to answer
the question:

{\bf "Does the unification of [SU(5)]$^3\,\,$ or [SO(10)]$^3\,\,$,
SUSY or not SUSY, really exist near the Planck scale?"}

Recently P.Ramond and Fu-Sin Ling [15] have considered
$[SO(10)]^3$ group of symmetry and have shown that it explains the
observed hierarchies of fermion masses and mixings.

{\Large \bf Acknowledgements:}
This investigation was supported by the grant RFBR 02-02-17379.

\end{document}